\documentclass[conference]{IEEEtran}

\usepackage[utf8]{inputenc}

\usepackage{academicons}
\usepackage{xcolor}

\newcommand{\orcid}[1]{\href{https://orcid.org/#1}{\textcolor[HTML]{A6CE39}{\aiOrcid}}}

\usepackage{enumerate}
\usepackage[shortlabels]{enumitem}
\usepackage{comment}
\usepackage{xspace}
\usepackage{graphicx}
\usepackage{multirow}
\usepackage{xurl}
\usepackage{hyperref}
\usepackage{balance}
\usepackage{amsmath}
\usepackage{tabularx}
\usepackage{ragged2e}
\usepackage{soul}
\usepackage{caption}
\usepackage{subcaption}
\usepackage{float}
\usepackage{wrapfig}

\usepackage{longtable}
\usepackage{paracol}
\usepackage{array}
\usepackage{tikz}
\newcommand*\circled[1]{\tikz[baseline=(char.base)]{
            \node[shape=circle,draw,inner sep=.8pt] (char) {#1};}}

\usepackage[]{graphicx}
\usepackage{duckuments}

\def\BibTeX{{\rm B\kern-.05em{\sc i\kern-.025em b}\kern-.08em
    T\kern-.1667em\lower.7ex\hbox{E}\kern-.125emX}}
\newcolumntype{L}{>{\raggedright\arraybackslash}X}

\makeatletter
\newcommand{\linebreakand}{%
  \end{@IEEEauthorhalign}
  \hfill\mbox{}\par
  \mbox{}\hfill\begin{@IEEEauthorhalign}
}
\makeatother

\begin{document}
\title{Virtualization-based Penetration Testing Study for Detecting Accessibility Abuse Vulnerabilities in Banking Apps in East and Southeast Asia}

 \author{
     \IEEEauthorblockN{
        Wei Minn\IEEEauthorrefmark{1}, 
        Phong Phan\IEEEauthorrefmark{2}, 
        Vikas K. Malviya\IEEEauthorrefmark{3}, 
        Benjamin Adolphi\IEEEauthorrefmark{4}, 
        Yan Naing Tun\IEEEauthorrefmark{1}, \\ 
        Henning Benzon Treichl\IEEEauthorrefmark{5},
        Albert Ching\IEEEauthorrefmark{2}, 
        Lwin Khin Shar\IEEEauthorrefmark{1}, 
        David Lo\IEEEauthorrefmark{1}
    }
     
     \IEEEauthorblockA{
        \IEEEauthorrefmark{1} Singapore Management University, Singapore\\
        \IEEEauthorrefmark{3}MIE-SPPU Institute of Higher Education, Qatar\\
        \IEEEauthorrefmark{2}i-Sprint Innovations Pte. Ltd, Singapore\\
        \IEEEauthorrefmark{4}Promon, Germany\\
        \IEEEauthorrefmark{5}Promon, Norway\\
     }
 }

\thispagestyle{plain}
\pagestyle{plain}

\maketitle

\begin{abstract}
Android banking applications have revolutionized financial management by allowing users to perform various financial activities through mobile devices. However, this convenience has attracted cybercriminals who exploit security vulnerabilities to access sensitive financial data. FjordPhantom,
a malware identified by our industry collaborator, 
uses virtualization and hooking to bypass the detection of malicious accessibility services, allowing it to conduct keylogging, screen scraping, and unauthorized data access. This malware primarily affects banking and finance apps across East and Southeast Asia region where our industry partner's clients are primarily based in. It requires users to be deceived into installing a secondary malicious component and activating a malicious accessibility service. 
In our study, we conducted an empirical study on the susceptibility of banking apps in the region to FjordPhantom, analyzed the effectiveness of protective measures currently implemented in those apps, and discussed ways to detect and prevent such attacks by identifying and mitigating the vulnerabilities exploited by this malware.

\end{abstract}


\section{Introduction}
\label{sec:intro}
Android applications have reshaped the banking landscape. Users can now conduct diverse financial activities through a simple tap on their smartphones or tablets. Furthermore, these apps also provide advanced functionalities such as budgeting tools, expense tracking, and tailored recommendations, which help users make well-informed financial choices and control their finances.
Security issues have also emerged with these substantial changes in the banking sector. The popularity of these apps has also drawn cybercriminals who aim to make money by taking advantage of security flaws in the app. Due to the sensitive nature of financial data, strict security measures are required to guard against fraud and unauthorized access.

FjordPhantom \cite{promon_fjordphantom} is one such malware that utilizes virtualization to target Android applications. It was identified by 
Promon \cite{promon_inc},
a Cybersecurity firm
and our industry partner.
FjordPhantom uses virtualization to disable the detection of malicious accessibility services. This allows it to operate without being flagged by security software.
FjordPhantom operates in two main stages. 
The first stage involves deceiving users into installing an app that seems benign. 
This app contains the code to disable the detection mechanisms for accessibility services. 
This is often done through social engineering tactics, such as fake updates or disguised apps. Attackers must also convince the user to activate the malicious accessibility service, which grants the service significant permissions.
In the second stage of the attack, the FjordPhantom takes advantage of the disabled detection to perform malicious activities such as keylogging, screen scraping, or other forms of unauthorized access to the app's data.

Motivated by the need to protect the clients of our industry partner from the threat of FjordPhantom in the mobile banking domain, we evaluate banking Android apps and report our findings in this paper. More specifically, we selected banking apps from the East and Southeast Asia where the majority of the clients of our industry partner are based in. 
Then, we developed a penetration testing system for exploring vulnerabilities of these apps, and analyzed the results of the attacks. To ensure the testing incurs no infringement of privacy or any form of damage to any person or organization, we put strict limits on the scope of the attack to local device only, and involves only a rudimentary disabling of accessibility service detection in the apps. Though limited in scope, the penetration testing adequately demonstrates the apps' vulnerability to FjordPhantom while avoiding any ethical implications. We also analyzed different protection measures that are used by the banking industry in their apps and checked their effectiveness against FjordPhantom. We then discussed adequate preventive measures to detect and patch the vulnerabilities that allow malware like FjordPhantom to successfully attack the app. Lastly, we discusses how these measures can be incorporated into prevailing software development practices in the industry.

The remainder of the paper is structured as follows. 
Section~\ref{attack} describes FjordPhantom -- its characteristics and how it works. 
Section~\ref{experiment} details the experiment setup to evaluate vulnerability of selected banking apps and protections they deploy.
Section~\ref{results} discusses how banking apps can be vulnerable against FjordPhantom and the current state of the distributions of those vulnerable apps, and discusses the insights drawn from this study and the implications for practitioners. 
Section~\ref{mitigation} presents how FjordPhantom can be prevented. 
Section~\ref{related_work} provides a summary of the related work. 
Section~\ref{conclusion} provides concluding remarks and future work.

\section{FjordPhantom Attack}  \label{attack}

\begin{figure}[h]
  \centering
  \includegraphics[trim=0.5cm .5cm 0.5cm .5cm,clip=true,width=.8\linewidth]{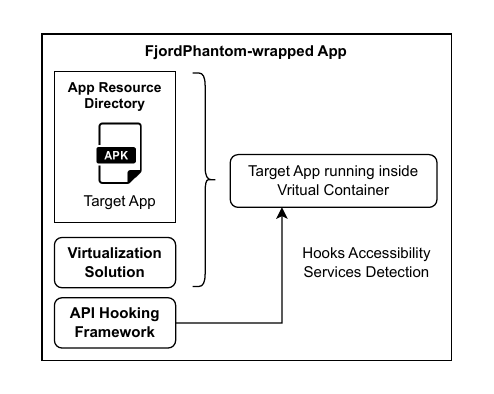}
  \caption{FjordPhantom Attack}
  \label{fig:FjordPhantom}
\end{figure}

The goal of the creators of FjordPhantom is to perform attacks based on malicious accessibility services (e.g., to obtain user input from the app, perform overlay attacks, or inject touch events) on a series of banking apps. However, since this is a very common attack performed by malware, banking apps typically detect malicious accessibility services and either warn the user not to proceed to even disallow the usage of the application altogether. Because of that, the creators of FjordPhantom needed to find a way to work around these protections. 
Typically, this is done using a repackaging attack where the logic to detect accessibility services is removed statically and the application is recompiled. This does, however, violate the integrity of the application and is commonly detected by anti-tampering measures already deployed inside banking apps. 
The authors of FjordPhantom therefore bypass the detection of malicious accessibility services without tampering the app by using a virtualization solution that enables them to run the target app in a virtualized environment that they control. In that environment, they can place runtime hooks into the functionality that is responsible for detecting malicious accessibility services without needing to modify the application, thus circumventing the anti-tampering mechanisms. 
As shown in Figure \ref{fig:FjordPhantom}, FjordPhantom relies on various components for attacking apps discussed below. %

\begin{figure}[h]
  \centering
  \includegraphics[trim=0.6cm 0.6cm 0.6cm .4cm,clip=true,width=.7\linewidth]{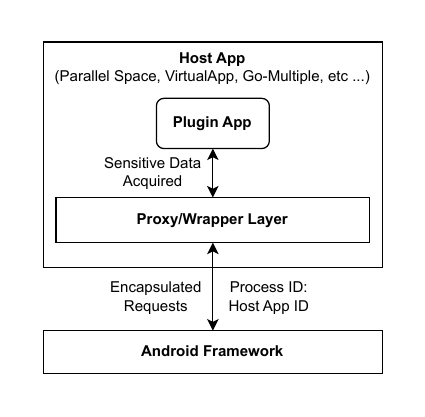}
  \caption{Broken Sandbox due to App Virtualization}
  \label{fig:virtualization}
\end{figure}

\textbf{App Virtualization.}
FjordPhantom makes use of the application virtualization solution which is an app (host) that hosts different apps (plugins) to run in the context of the host app. To achieve this, application virtualization solutions encapsulate the system requests of the plug-in applications and act as the proxy between the plug-ins and the Android system~\cite{dai_2020}. This breaks the Sandbox security feature that restricts different apps and services from accessing each other's file system and memory because the plugins' requests are encapsulated by the host's process ID. Figure~\ref{fig:virtualization} depicts an example permission escalation scenario where a plugin can access sensitive data after inheriting permissions given to the host app by the Android system.
Exploiting this concept of encapsulation and sharing the same process ID, any app (like FjordPhantom) that hosts plugins (and uses the virtualization solution) have access to the memory and file systems of the plugin apps.

\textbf{API Hooking.} Next, FjordPhantom uses a Java API hooking framework to intercept and modify API communication between the target app and the Android system during runtime. Hooking allows FjordPhantom to avoid having to statically modify the original code which would trigger anti-tampering hardening measures of the target app~\cite{hooking}. 
More specifically, FjordPhantom 
intercepts communications with Google Play Services to make it seem as unavailable. Security features such as SafetyNet that detect rooting uses Google Play Services, and thus apps using SafetyNet might be tricked into skipping the root detection part during runtime~\cite{playintegrity}.

\textbf{Repackaging.} The target app, together with the virtualization and hooking solutions, are repackaged into a new app whose appearance and usability resemble the target app in the eyes of the victim. The repackaged app handles the logic of loading the target app's APK and invocations of virtualization and hooking solutions.

\textbf{Accessibility Abuse.}
Accessibility services are granted broad permissions when enabled by users, which allow them to interact with the interface elements of any app on the device, including reading content on the screen and injecting inputs like clicks and keystrokes~\cite{xudva}. Usually, malicious apps will be prevented by the Sandbox mechanism from interacting with or accessing resources of other processes. 
However, FjordPhantom uses virtualization to break the Sandbox, hooks the target app to bypass accessibilility service detection measures, and invoke the accessibility services to steal information from other apps.
This can lead to not only passive attacks like screen recording and touch logging, but also active attacks like auto clicking-driven attacks~\cite{autoclicking}.

\section{Experiment Design} 
\label{experiment}

\begin{table}[t]
\caption{Evaluated Banking Apps}
    \centering
  \label{tab:appsbycountry}
  \begin{tabular}{|c|c|r|}
    \hline
     \textbf{Country} & \textbf{\# Apps Analyzed} & \textbf{Total Downloads} \\ \hline
    Thailand &  14 & $49,650,000$+\\
    \hline
    Japan & 13 & $40,300,000$+ \\
    \hline
    Malaysia & 12 & $5,500,000$+\\
    \hline
    Indonesia & 24 & $168,600,000$+ \\
    \hline
    Korea & 10 & $117,000,000$+\\
    \hline
    Singapore &  8 & $23,200,000$+ \\
    \hline
    Hong Kong & 9 & $1,700,000$+ \\
    \hline
    \textbf{Total} & 83 & $405,950,000$+ \\
    \hline
  \end{tabular}
\end{table}

\subsection{Banking Apps Evaluated}
We conducted penetration testing of FjordPhantom on banking and finance apps from seven countries from East and Southeast Asia where most of the clients of our industry partner are based in. In total, we evaluated 83 apps with collective downloads of more than 405 million. 
We reviewed similar works by Chen et al. \cite{chen_2020}, and Dai et al. \cite{dai_2020} to ensure safety, ethics and privacy in this experiment due 
to the highly sensitive nature of financial data. 
Table \ref{tab:appsbycountry} shows the number of banking apps, their respective countries, and the number of downloads. We chose the apps from the Google Play store as Google Play is the largest app store in the world and provides convenient access to app versions that come directly from official app developers of the banks without the risk of tampering.

\subsection{Static Analysis} \label{sec:static}

As we expect the apps to come with security mechanisms to protect apps from reverse engineering, tampering, and other malicious activities, we seek more insight into the distribution of such measures by way of static analysis. We use APKiD which is a popular\footnote{APKiD has 2.3K+ GitHub stars as of August 15, 2025.} tool to detect if an APK is shielded by Runtime Application Self-Protection (RASP) solutions~\cite{apkid}. The RASP solutions are of 3 types: packers, protectors, and obfuscators. Packers bundle the app’s code and resources into a single, encrypted, or compressed, package which is only unpacked in memory during runtime to make static analysis harder;
protectors focus on securing the app through a combination of obfuscation, encryption, anti-tampering, and anti-debugging techniques;
obfuscators transform the code into a less readable form
by renaming variables, methods, and classes to meaningless names, control flow obfuscation, string encryption, inline expansion, dummy code insertion, etc.  
In addition, APKiD also reports if an APK contains standalone hardening measures, namely anti-disassembly, anti-debugging, and anti-virtual machine (anti-VM). We shall refer to these three methods as \emph{hardening measures}. 

\subsection{Penetration Test Setup} \label{sec:pentest}

\begin{figure}[t]
  \centering
  \includegraphics[trim=0.cm 0.64cm 0.cm 0.65cm,clip=true,width=.9\linewidth]{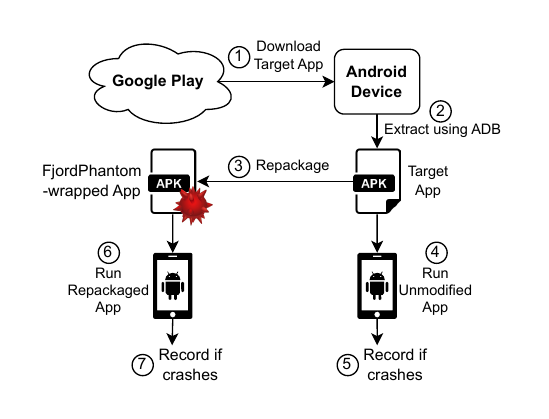}
  \caption{Penetration Testing Flow}
  \label{fig:pentest}
   \vspace{-1.5em}
\end{figure}

We have created an automated system (Figure~\ref{fig:pentest}) for validating whether a given app is vulnerable to virtualization attacks like FjordPhantom. It comprises a penetration test which is performed under strictly-defined attack scope (bypass the first dialog box shown on launch that alerts the user about potentially harmful enabled accessibility service, and terminate immediately) to prevent any form of damage caused to any person or organization.
When started, the system receives a list of applications as input. Then it performs the following steps for each app:

First, the application is downloaded from Google Play on an Android device (\circled{1}). 
Our system launches the Google Play app and instructs it to show the screen for downloading the target app. This is done using the \texttt{am} command shipped with all Android versions. 
The system then simulates a click on the install button in the Google Play app. For that, it first needs to find the location of the button. This is achieved using the \texttt{uiautomator} command, which dumps the UI hierarchy of the device. The system can then simulate a tap on the button using the \texttt{input} command.
After the app has been installed, it is extracted from the device it was installed on. This is achieved using \texttt{adb} (\circled{2}).
Once the app is downloaded, it is repackaged to perform the virtualization-based attack (\circled{3}). The system bases the attack on the FjordPhantom as described in Section~\ref{attack}. This repackaged app runs the APK file of the target app that is embedded into its resource folder in a virtual container and applies hooks to the accessibility services detection.

\begin{figure*}
\captionsetup{justification=centering}
  \begin{minipage}[t]{.32\linewidth}
  \centering
      \includegraphics[trim=1.1cm .45cm .7cm 0.6cm,clip=true,width=\linewidth]{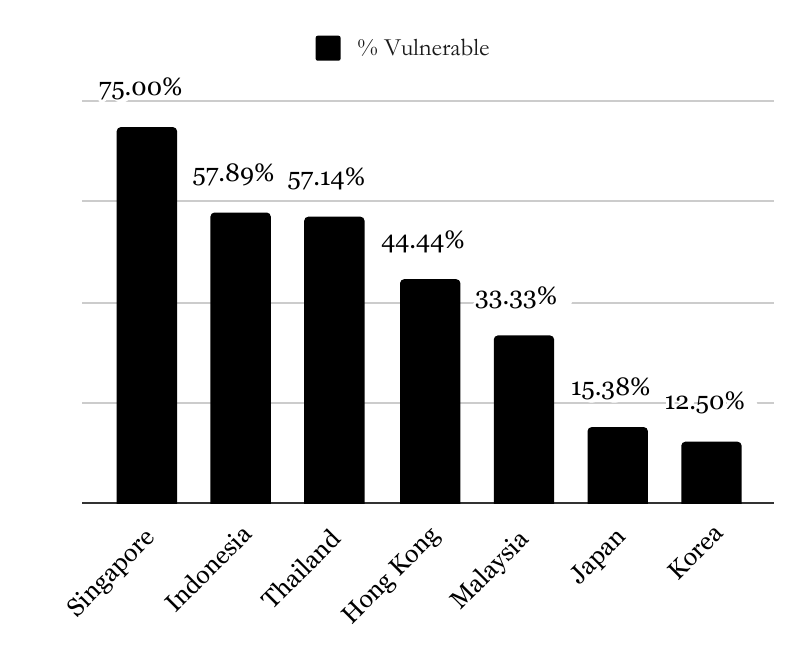}
  \caption{Percentage of Apps Vulnerable to FjordPhantom Attacks by Country}
  \label{fig:percentagevulnerable}
  \end{minipage}\hfil
  \begin{minipage}[t]{.32\linewidth}
  \centering
  \includegraphics[trim=.8cm 0.5cm 1cm 0.6cm,clip=true,width=\linewidth]{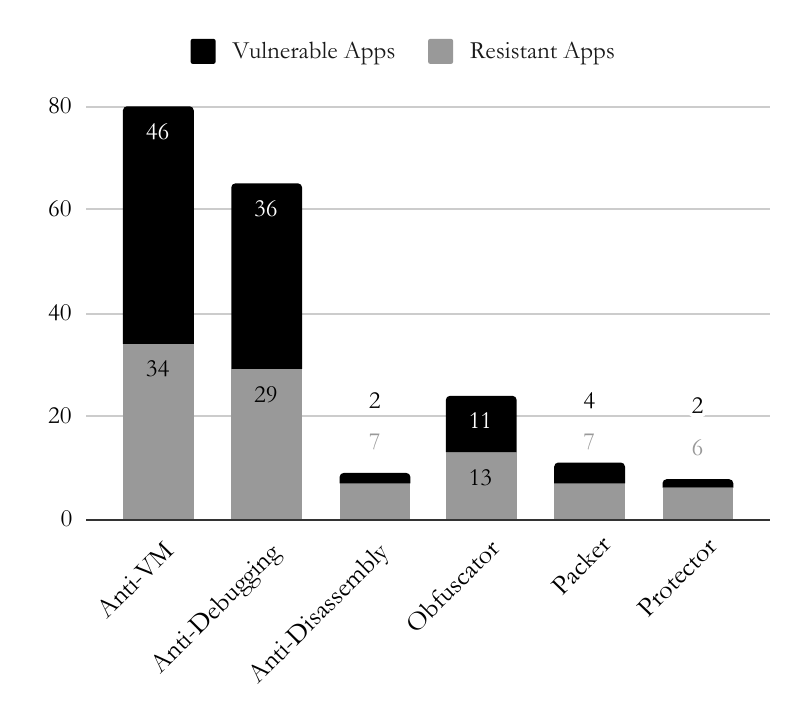}
  \caption{Distribution of Hardening Measures and RASP Solutions}
   \vspace{-.8em}
  \label{fig:distributionprotection}
  \end{minipage}\hfil
  \begin{minipage}[t]{.32\linewidth}
  \centering
  \includegraphics[trim=1.5cm 1.1cm 0.9cm 0.6cm,clip=true,width=\linewidth]{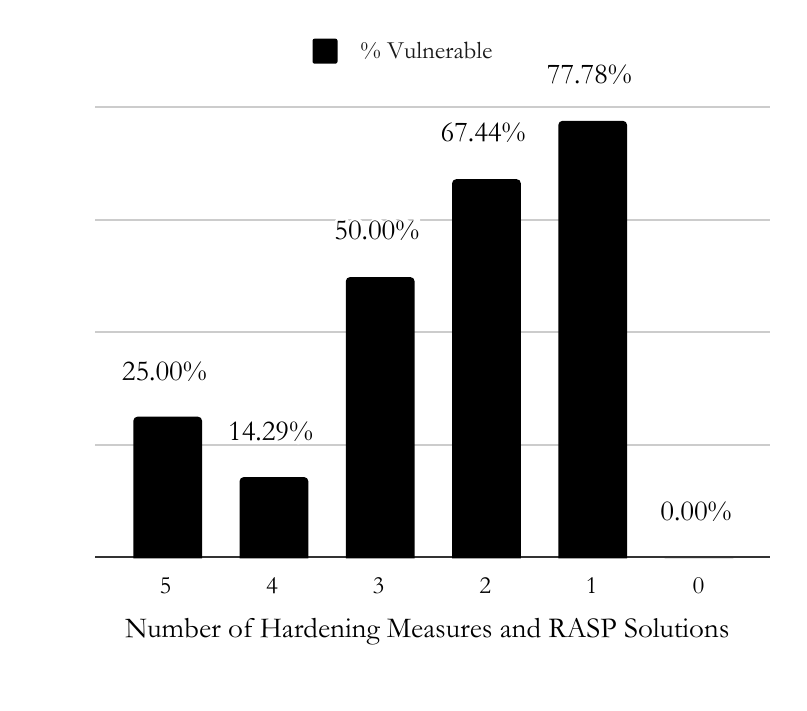}
  \caption{Number of Hardening and RASP Combined vs. Vulnerability}
   \vspace{-.8em}
  \centering
  \label{fig:numberprotections}
  \end{minipage}%
\end{figure*}

To make sure that there are no problems with running the app in our test environment, we first install the unmodified version of the app and make sure that it does not crash by running the app and then observing the UI hierarchy of the device to make sure that the application stays in the foreground for 30 seconds (\circled{4}). Since Android tends to re-launch apps that crash quickly after startup, the system also makes sure that the process ID of the app does not change during that period (\circled{5}).

If the system finds the unmodified version running correctly, it repeats the same procedure with the repackaged version of the app (\circled{6}). We assume that if an app runs in a virtual environment like this, it should not run properly because if it continues running, it could be manipulated (\circled{7}). Anti-VM logic, for example, will try to warn the user that it is running in a virtual environment, but the environment could just prevent the message from being shown to the user. Because of that, our system assumes that if the unmodified version of the app runs without problems and the version of the app that runs in a virtual environment crashes, the application is assumed to be not vulnerable to FjordPhantom. Conversely, if the unmodified version of the app runs without problems and the version of the app that runs in a virtual environment does not crash, the application is assumed to be vulnerable to FjordPhantom.

\section{Results and Discussions} \label{results}

Figure~\ref{fig:percentagevulnerable} shows the percentage of banking apps that are vulnerable to FjordPhantom in the penetration test for each of the selected 7 Asian countries. Figure~\ref{fig:distributionprotection} depicts the distribution of protection measures detected by APKiD and the percentage of apps that are vulnerable to or resistant to FjordPhantom attacks concerning each protection.
Figure~\ref{fig:numberprotections} shows the percentage of vulnerable apps for each combined number of hardening measures and protection solutions.

Figure~\ref{fig:percentagevulnerable} shows that $75\%$ (6 out of 8) of the banking apps in Singapore that we analyzed are vulnerable against FjordPhantom. This is true for the majority of banking apps in Indonesia ($57.89\%$) and Thailand ($57.14\%$) we analyzed as well. In contrast, Japan ($15.38\%$) and Korea ($12.50\%$) have the lowest rate of vulnerability against FjordPhantom. In total, our experiments show that $43.37\%$ (36 out of 83) of the banking apps across the 7 Asian countries are vulnerable. \textbf{The implication is that approximately 225 million Android users (based on cumulative downloads of those 36 apps) in East and Southeast Asia are subject to accessibility abuse attacks like FjordPhantom.} This may explain the sharp rise of banking-related phishing scams in Southeast Asia~\cite{bankinfo}.

From Figure~\ref{fig:distributionprotection}, we observe that anti-VM is the most widely implemented hardening measure amongst all the banking apps analyzed (80 out of 83 apps), while anti-disassembly is the least implemented (9 out of 83 apps). Interestingly, the majority of the apps ($78\%$) with anti-disassembly measures are resistant to FjordPhantom attacks. It seems to be effective even when FjordPhantom does not need access to the original code for static modification. On the other hand, $57.5\%$ (46 out of 80 apps) of apps with anti-VM are vulnerable to FjordPhantom attacks even though FjordPhantom uses a virtualization solution to load the target app into a shared context with FjordPhantom. 
\textbf{Majority of apps (6 out of 8 apps) that employ protector-type RASP solutions are resistant to FjordPhantom but few apps (only $9\%$ of total apps) employ such protector solutions}. In our future work, we plan to dive deeper into the implementation details of hardening measures and RASP solutions, and more systematically investigate their effectiveness against accessibility abuse attacks like FjordPhantom.

In Figure~\ref{fig:numberprotections}, we take a look at the combined number of different hardening measures and RASP solutions implemented in the app evaluated, and how it affects the app's resistance to FjordPhantom. We found that the highest number of different hardening and protection solutions employed in an app we analyzed is five. \textbf{In general, we see a trend of increasing vulnerability as we decrease the number of hardening measures and RASP solutions employed.} There are two apps in our evaluation for which APKiD did not detect any hardening measure or RASP solution and yet, they are not vulnerable to FjordPhantom. We attribute this observation to the particular characteristics of those apps and the possible implementations of custom hardening measures that APKiD does not recognize. We plan to further investigate this in future work.

We have reported our findings to the stakeholders from the banking industry and Government Agencies by conducting a seminar~\cite{seminar}.

\section{Mitigation} \label{mitigation}

\begin{figure}[t]
  \centering
  \includegraphics[trim=0.5cm 0.6cm 0.5cm 0.3cm,clip=true,width=.95\linewidth]{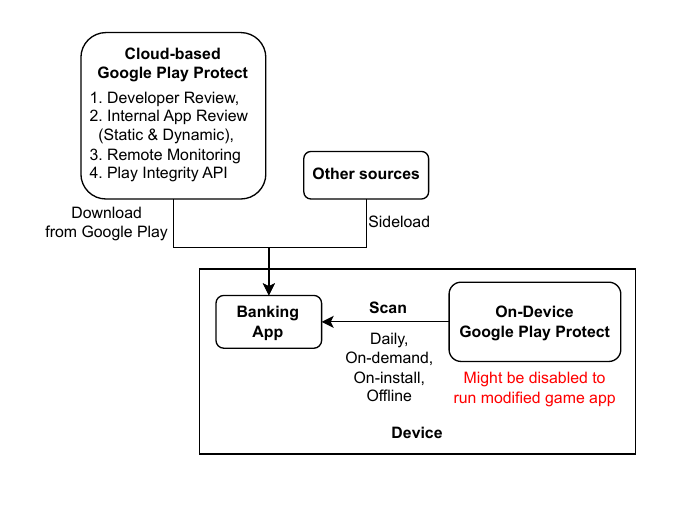}
  \caption{Google Play Protect's App Scanning Process}
  \label{fig:automatedhardening}
  \vspace{-.75em}
\end{figure}

\subsection{Automated Detection by Google Play Protect}

As all the apps we evaluate are downloaded from Google Play, we focus on Google Play Protect (GPP) which offers protections against malware and Potentially Harmful Applications (PHAs) on two levels: cloud-based and on-device~\cite{gpp}, as depicted in Figure~\ref{fig:automatedhardening}. 
On Cloud, GPP performs static and dynamic analysis on the applications and
also collects data about unknown apps inside users' devices, and reports by security researchers, to build up a comprehensive database of malware and PHAs. Locally, GPP does scanning of applications at regular intervals, on-demand, or on-installation - online or offline. 

FjordPhantom may still slip through the detection mechanisms of GPP as it is a relatively unknown attack at the time of our experiments~\cite{FjordPhantom_mitre}. Therefore, we believe that its malware signatures have not been registered in the GPP database yet.
As FjordPhantom is a virtualization-driven malware, organizations can set policy for their mobile devices to not allow invocations of virtualization frameworks, and dynamic DEX/APK loading by apps. Users may only download their banking applications from the official source, Google Playstore, to avoid the possibility of encountering a FjordPhantom-wrapped app. These are stop-gap solutions while the signature of FjordPhantom is being processed into the GPP malware signature database for detection on future installations.

However, even if GPP can detect and prevent the installation and execution of FjordPhantom-wrapped apps, there may be certain cases where the users deliberately disable GPP to install and run their own repackaged apps which would otherwise be blocked by GPP. This is common when users try to circumvent the anti-cheat mechanisms of mobile games, and DRM measures of other apps like streaming services. In this case, on-device GPP would not be able to stop the execution of the FjordPhantom-wrapped apps, thus the need for app hardening measures that do not rely on the existence of external protection services.

\subsection{Automated Hardening}

\begin{figure}[t]
  \centering
\captionsetup{justification=centering}
  \includegraphics[trim=0.5cm 0.75cm 0.5cm .75cm,clip=true,width=.8\linewidth]{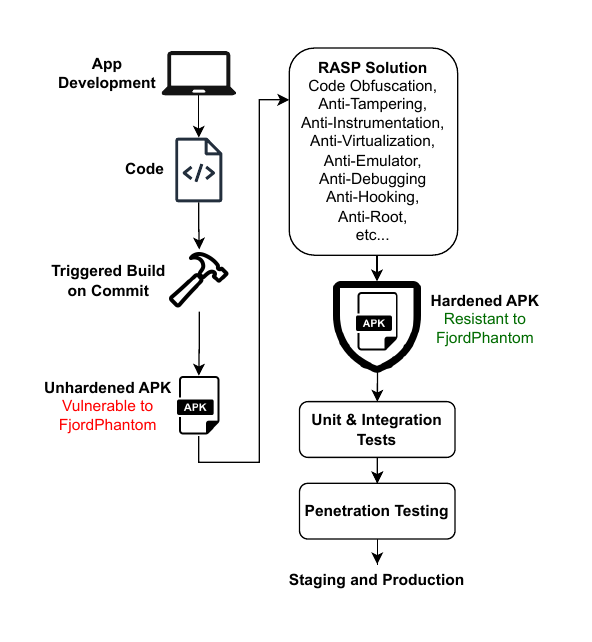}
  \caption{Automated App Hardening Process Integrated into CI/CD Pipeline}
  \label{fig:CICD}
   \vspace{-0.75em}
\end{figure}

Protecting apps against a wide range of attacks can be challenging for organizations with resource and manpower constraints~\cite{thomas_security}.
To solve this problem, different Cybersecurity companies - including our industry collaborator, Promon~\cite{promon_inc}, offer services in the form of RASP solutions that automatically add protection to apps without the app developer needing to be an expert in application hardening~\cite{haupert_honey}. 
Promon SHIELD is one such solution that assists developers in making their apps secure \cite{promon_app_shield}. 
This shield is different from other security solutions which test applications for different types of attacks in the way it modifies the application's binary codes and makes it prone to intrusion, tampering, reverse engineering, and malware attacks.

Once the developer has coded up and built an app, its APK file is submitted to an automatic hardening service by our industry collaborator, which applies various protections based on a provided configuration file. 
Figure \ref{fig:CICD} shows details of how Promon SHIELD can be integrated into the CI/CD pipeline which is widely practiced in the industry for streamlined development and deployment~\cite{cicd}. While developing an app, once the development phase is done and the code is committed, the automated build is triggered and generates an APK file that is functional enough to be installed on the end user's device. However, this APK file is unhardened and can be prone to FjordPhantom. In the next stage of the pipeline, Promon SHIELD automatically modifies the APK file on binary level and
adds protective measures into the application by injecting byte-codes related to obfuscation, tamper detection, anti-instrumentation, anti-virtualization, anti-emulation, anti-debugging, anti-hooking, anti-root and et cetera. Since the binary of the APK has been modified, the functionalities of the now-shielded APK need to be verified by unit and integration tests and then penetration testing for which setup we presented in Section~\ref{sec:pentest}. Finally, the app becomes ready for production. This way, we can integrate Promon SHIELD to make the app readily self-protecting against FjordPhantom malware in whichever environment it is released into.

\section{Related Work} \label{related_work}

\hspace{\parindent}\textbf{Virtualization-based Malware.}
Zhang et al. \cite{zhang_2019} examined the security risks of app virtualization in Android. They analyzed 32 popular frameworks, identifying seven attack vectors and vulnerabilities to malware distribution.
Dai et al. \cite{dai_2020} studied app-level virtualization on Android and related security threats. Analyzing over 160 apps from popular markets, they identified attack vectors like privilege escalation and reference hijacking. 
Shi et al. \cite{shi_2020} 
proposed VAHunt which identifies app-virtualization engines in APK files and uses data flow analysis to differentiate between benign and malicious loading strategies. 
These research works concentrate on detecting virtualization-based attacks. Our work goes further by not only analyzing this kind of attack, but also evaluating various protections measures, and discussing a solution by our industry collaborator that can be integrated into the CI/CD pipeline.

\textbf{Android Banking App Security.}
Chen et al.~\cite{chen_2020} conducted large-scale empirical study of Android banking apps and identified their vulnerabilities on 4 levels: input, data storage, data transmission, and communication infrastructure, and their distributions across different countries and regions.
Sharma et al.~\cite{sharma_2023} provided a general overview of common types of malware in banking apps such as trojans, spyware, and keyloggers. They also presented practices such as Application Provenance, and Asset Based Access Control to ensure the confidentiality and integrity of sensitive user data.
Malviya et al.~\cite{malviya_industry} performed evaluated vulnerabilities in 28 banking apps from 10 countries by performing 11 attacks, and discussed preventative measures to identify and defend against those attacks.
Jung et al.~\cite{Jung2013} discussed repackaging attacks that remove specific parts of the code for hardening measures in Android banking apps, and its countermeasures such as code obfuscation and attestation.
To our knowledge, we are the first to cover specifically virtualization-related vulnerabilities in Android banking apps, and more specifically, in the region of East and Southeast Asia.

\section{Conclusion and Future Work} \label{conclusion}

In this paper, we discussed FjordPhantom, a virtualization-based accessibility abuse malware, that targets apps in the banking and financial sector. 
We conducted penetration testing of these apps and exploited their vulnerabilities against FjordPhantom.
We also analyzed hardening measures and protection solutions implemented in these apps and their efficacy in protecting against FjordPhantom. We found that over 225M Android users are subject to FjordPhantom. 
We discussed details on how 
Promon SHIELD, a solution that protects against FjordPhantom,
can be integrated into the CI/CD pipeline that is commonly practiced in the industry.
In the future, we plan to explore supposed peculiarities in the implementation of apps that defied our expectations during penetration testing, and detailed empirical evaluation of the efficacy of our industry collaborator's and other RASP solutions against different forms of attacks.
Overall, this study provides insights into how the banking industry applies security concepts such as hardening measures, RASP solutions, and penetration testing to bridge the gap between the technological edge of attackers and protections for mobile banking apps.

\balance
\bibliographystyle{IEEEtranS}
\bibliography{references}

\end{document}